\documentclass[revsymb,prb,twocolumn]{revtex4}%
\usepackage{graphicx}
\usepackage{bm}
\usepackage{slashed}
\usepackage{amsmath}
\usepackage{amsfonts}
\usepackage{wasysym}
\usepackage{amssymb}%
\providecommand{\U}[1]{\protect\rule{.1in}{.1in}}
\providecommand{\U}[1]{\protect\rule{.1in}{.1in}}
\newcommand{\be}{\begin{equation}}
\newcommand{\en}{\end{equation}}

\begin{document}
\title{AC magnetic response of highly nonlinear soliton lattice in a monoaxial chiral helimagnet}
\author{J. Kishine$^{1,2}$ and A.S. Ovchinnikov$^{3,4}$}
\affiliation{$^{1}$ Division of Natural and Environmental Sciences, The Open University of
Japan, Chiba 261-8586, Japan}
\affiliation{$^{2}$ Institute for Molecular Science, 38 Nishigo-Naka, Myodaiji, Okazaki,
444-8585, Japan}
\affiliation{$^{3}$ Institute of Natural Science and Mathematics, Ural Federal University,
Ekaterinburg 620002, Russia}
\affiliation{$^{4}$ Institute of Metal Physics, Ural Division, Russian Academy of Sciences,
Ekaterinburg 620219, Russia}
\date{\today }

\begin{abstract}
We present the theory of nonlinear ac magnetic response for the highly
nonlinear regime of the chiral soliton lattice. Increasing of the dc magnetic
field perpendicularly to the chiral axis results in crossover inside the
phase, when nearly isolated $2\pi$-kinks are partitioned by vast ferromagnetic
domains. Assuming that each of the kink reacts independently from the other
ones to the external ac field, we demonstrate that internal deformations of
such a kink give rise to the nonlinear response.

\end{abstract}

\pacs{Valid PACS appear here}
\maketitle

\section{Introduction}

Measurements of nonlinear ac magnetic response can provide extremely useful
information for understanding physical properties of various magnetic
compounds. This powerful magnetic diagnostics is conventionally used to
elucidate a dynamics of magnetic domains in ferromagnets \cite{Sato1981} in
addition to allowing the determination of the phase transition temperature
\cite{Hashimoto1973}. This experimental technique has applications in studies
of low-field magnetic hysteresis of ferromagnetic
\cite{Rayleigh1887,Milstein1975} and helimagnetic materials
\cite{Mito2010,Mito2015}, nonlinear response in ferroelectrics
\cite{Ishibashi1994} and molecule-based magnets with structural chirality
\cite{Mito2009,Mito2012}. Another broad research area involves studies of
nonlinear susceptibility in the vicinity of spin glass transition temperature
\cite{Suzuki1977, Miyako1979, Fujiki1981}. A chirality driven mechanism due to
fluctuations of a dynamical noncollinear spin order variable, which defined as
a vector product of two nearest-neighbor lattice spins, was proposed to
explain new universality classes of experimentally observed spin-glass
critical phenomena
\cite{Kawamura1992,Maleyev2002,Maleyev2004,Kawamura2010,Campbell2010}.

Recent investigations of chiral magnets have become a new landmark of the
nonlinear response applications. These materials comprise both the chirality
of critical fluctuations, such as in the case of MnSi
\cite{Grigoriev2005,Papas2011}, and the ac magnetic response of magnetic
domains found, for example, in Cr${}$Nb$_{3}$S${}_{6}$ \cite{Tsuruta2016}. In
the latter case, the phase diagram of the chiral helimagnet as a function of
temperature $T$ and a dc magnetic field $H_{\text{dc}}$ was constructed by
means of the ac magnetic susceptibility measurements. It was revealed that
when $H_{\text{dc}}=0$ the ac response consists of a giant third-order
harmonic component ($M_{3\omega}$) along with a first-order part ($M_{1\omega
}$) at the transition between the chiral helimagnetic (CHM) state to the
paramagnetic (PM) state. The non-zero dc field applied perpendicular to the
helical axis transforms the CHM state in favor of the chiral soliton lattice
(CSL) state. At small $H_{\text{dc}}$, ferromagnetic regions are poorly
expressed against the predominant magnetic helicoidal order. For this regime,
it was found that the $M_{3\omega}$ is drastically suppressed, i.e. the
transition between the CSL and the PM states is characterized only by a linear
magnetic response. At higher dc fields, the ferromagnetic domains start to
grow that allows the CSL state to be viewed as a regular arrangement of 2$\pi
$-kinks in the ferromagnetic background. With increasing temperature, this CSL
state is transformed into the forced ferromagnetic (FFM) state, and eventually
the PM state is reached. The transition between the CSL state and the FFM
states is again accompanied by a large $M_{3\omega}$ as well as a complex
$M_{1\omega}$, whereas the FFM-PM transition gives rise a linear magnetic
response without any energy loss. These data reported by Mito al.
\cite{Tsuruta2016} were supplemented by a comprehensive study made by Clements
at al. \cite{Clements2018}. They could trace the dc magnetic field dependence
of the ac magnetic response for the first five harmonic components and have
amply demonstrated a presence of the noticeable $M_{2\omega}$ component as the
ferromagnetic-domain-rich CSL evolves into the FFM state. In small ac-fields,
the $M_{2\omega}$ is connected to the breaking of the time inversion symmetry,
and the large signal reflects the presence of spontaneous magnetization
\cite{Hashimoto1973}. The $M_{3\omega}$, in contrast, is associated with the
microscopic breaking of spatial symmetry of magnetic moments \cite{Mito2009},
thereby confirming the presence of the ferromagnetic domains in the highly
nonlinear CSL state.

The peculiarities discovered in the nonlinear response are closely linked to
the $H-T$ phase diagram of the chiral helimagnet Cr${}$Nb$_{3}$S${}_{6}$,
detailed structure thereof is still actively debated. The nature of the
possible phase transitions has been addressed both experimentally and
theoretically. The dc magnetization and magnetic entropy change measurements
\cite{Ghimire2013,Clements2017} testify apparently a second-order phase
transition to the FFM state at magnetic fields above the critical field of the
incommensurate-commensurate (IC-C) phase transition. At moderate and low
magnetic fields the onset of the chiral IC phase was detected, including
crossover between the nonlinear and highly nonlinear regions of the CSL state
along with complementary crossover between the CHM and CSL states. These
investigations also confirm the existence of the weakly nonlinear CSL and the
concurrent disappearance of the CHM phase for small non-zero applied fields in
temperature region above and below the Curie temperature $T_{c}\sim$ 130.7 K.
A possibility of first-order behavior at the phase transition in small
magnetic fields were also argued. The metamagnetic crossover from the weakly
nonlinear to the highly nonlinear CSL regime has been verified by measurements
of magnetoresistance \cite{Togawa2013}. The theoretical studies by Laliena et
al. \cite{Laliena2016,Laliena2017} predict a transition line between the
highly nonlinear CSL and the FFM states at low temperature and high dc fields
as being of second-order (continuous) nucleation type transition
\cite{Gennes1975}. The boundary between the CHM and the PM state may be
categorized as the second-order instability type transition \cite{Izyumov1985}
according to de Gennes classification. The second-order line sections are
separated by a line of first-order transitions with two tricritical points as
a terminus at intermediate temperatures and magnetic fields. Experimental
validation of these tricritical points remains controversial \cite{Han2017}.
Another examples of the nucleation type transition are the transition at the
lower critical field of type-II superconductors and the cholesteric-to-nematic
\ transition in liquid crystals. The former analogy was discussed in Refs.
\cite{Shinozaki2019, Masaki2020}, where a similarity of surface barrier
between the monoaxial chiral helimagnet and type-II superconductors was
pointed out.

In our work we present the theory of nonlinear ac magnetic response for the
highly nonlinear regime of the chiral soliton lattice. The state emerging as a
result of crossover under increasing the dc magnetic field perpendicularly to
the chiral axis may be regarded as a regular arrangement of nearly isolated
$2\pi$ kinks partitioned by vast ferromagnetic domains. It is separated from
the FFM state by the nucleation type continuous transition and may be modeled
as particles that repel each other by a force which decay exponentially as a
distance between the kinks increases \cite{Rubinstein1970}. Then, it is
appropriate to assume that each of the kinks reacts independently from the
other ones to the external ac field. We argue that the appearance of
high-order harmonics and phase shifts is related to quasi-localized
excitations triggered by internal deformations of the separate kink.

To provide support for the picture presented above we formulate a model of the
magnetic soliton lattice relevant for the chiral helimagnet Cr${}$Nb$_{3}$%
S${}_{6}$ and explain how crossover between the weakly and highly nonlinear
CSL regimes originates from Fourier decomposition of the CSL configuration.
Furthermore, we find spectrum of Gaussian fluctuations of a single kink,
thereby specifying its internal deformations. Using the spectrum, a Lagrangian
formalism based on the collective coordinate method
\cite{Book2015,Kishine2012,Kishine2016} is developed to describe kink dynamics
driven by an external ac magnetic field. Solving of the corresponding
dynamical equations is reduced to a challenge how to get periodic (Floquet)
solutions. We apply the algorithm elaborated by Erugin \cite{Erugin} to
successfully overcome the problem and use the periodic solutions to recover
higher-order harmonic components of magnetization together with related phase
shifts. Our analysis reveals that the order parameter which characterizes the
second order phase transition of the nucleation type, namely a density of
kinks, has a crucial role to play in hierarchy of these higher-order harmonic
components. This allows to establish the limits of our theory and predicts the
onset of the linear response regime while approaching the FFM phase boundary.

This paper is organized as follows. In Sec. II, we describe the model and
summarize key details of the ground state and the excitations of the highly
nonlinear CSL state. Here, the Lagrangian formalism to describe dynamics of a
single kink is presented. In Sec. III, the periodic solution of the dynamical
equations is looked for which is used to derive higher order harmonics of
nonlinear magnetic response. The conclusions are given in Sec. IV.

\section{Model}

The layered structure of Cr${}$Nb$_{3}$S${}_{6}$ consists of 2H-type planar
NbS${}_{2}$ with the Cr atoms intercalated between the planes and belongs to
the non-centrosymmetric hexagonal space group P6${}_{3}$22 \cite{Togawa2016}.
The localized moments of the Cr${}^{3+}$ ions (the spin $S=3/2$) are oriented
in the crystallographic ab plane and exhibit strong single-ion
anisotropy\cite{Moriya1982,Miyadai1983}. As has been repeatedly proven the
qusi-1D model of the chiral soliton lattice \cite{Dzyaloshinskii1964} explains
property of the compound beautifully that was amply confirmed by Togawa et al.
via the Lorentz microscopy experiments \cite{Togawa2012}.

In the continuum approximation, the monoaxial chiral helimagnet is described
by the Hamiltonian, $H=\int dz\mathcal{H}$, with the density
\begin{align}
\mathcal{H}  &  =\frac{JS^{2}a_{0}}{2}\left[  \left(  \partial_{z}%
\theta\right)  ^{2}+\sin^{2}\theta\left(  \partial_{z}{\varphi}\right)
^{2}\right] \nonumber\\
&  -DS^{2}\sin^{2}\theta\left(  \partial_{z}{\varphi}\right)  -HSa_{0}%
^{-1}\sin\theta\cos{\varphi}, \label{continuumH}%
\end{align}
where $a_{0}$ is the lattice constant and the semiclassical spin
\[
\boldsymbol{S}\left(  z\right)  =S\left[  \sin\theta\left(  z\right)
\cos{\varphi\left(  z\right)  },\sin\theta\left(  z\right)  \sin
{\varphi\left(  z\right)  },\cos\theta\left(  z\right)  \right]
\]
is specified by the polar coordinates. Here, $J>0$ is the strength of the
nearest-neighbor ferromagnetic exchange coupling. The mono-axial DM vector
$\boldsymbol{D}=D\hat{\boldsymbol{e}}_{z}$ directed along the $z$-axis is
parametrized by the constant $D$. To stabilize the CSL state, a static
magnetic field $H$, measured in units $g\mu_{B}$, is applied perpendicularly
to the $z$-axis.

The CSL ground state is given by $\theta_{0}=\pi/2$ and $\varphi_{0}%
(z)=\pi+2\text{am}(\bar{z})$, where the dimensionless coordinate $\bar
{z}=(m/\kappa)z$ is introduced with $m^{2}=H/(a_{0}^{2}JS)$. The
$\text{am}(\ldots)$ is the Jacobi amplitude function depending on the elliptic
modulus $\kappa$ ($0\leq\kappa\leq1$). The modulus $\kappa$ is determined by
minimizing an energy that gives $\kappa=4mE/\pi q_{0}$, where $q_{0}=D/a_{0}%
J$, and the CSL spatial period, $L_{\text{CSL}}=2\kappa K/m$, with $K$ and $E$
being the elliptic integrals of the first and second kind, respectively.

\begin{figure}[ptb]
\begin{center}
\includegraphics[width=60mm,bb=0 0 1000 1200]{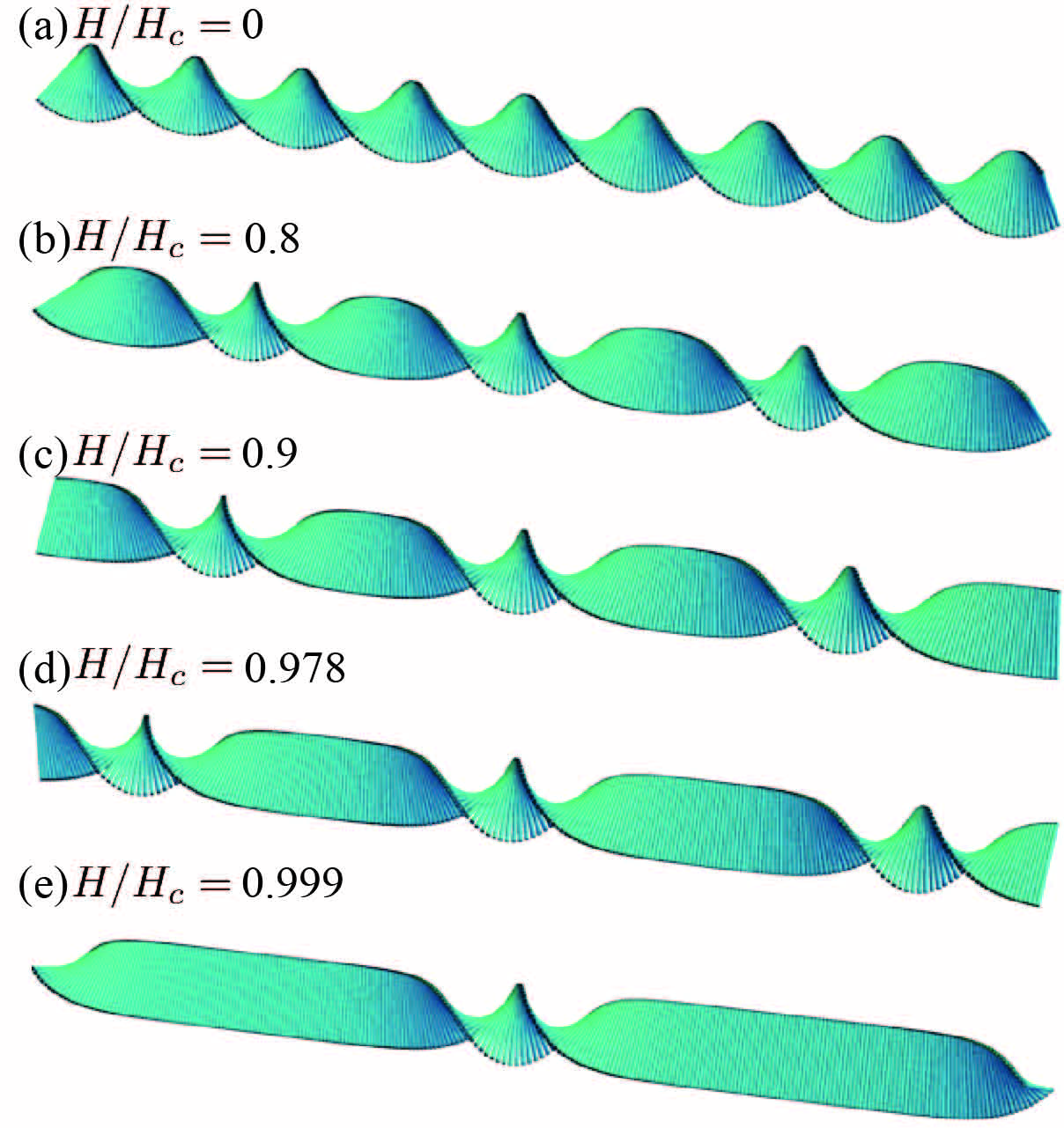}
\end{center}
\caption{Formation of the chiral soliton lattice for $H/H_{c}=0.8$(a),
$0.9$(b), $0.978$(c) and $0.999$(d).}%
\label{fig01}%
\end{figure}

The CSL undergoes a continuous transition to the forced ferromagnetic state
above the critical field
\begin{equation}
H_{\text{c}}=\left(  \frac{\pi q_{0}a_{0}}{4}\right)  ^{2}JS
\label{critical-field}%
\end{equation}
attained as $\kappa\rightarrow1$. The evolution of the CSL upon increasing $H$
is schematically shown in Fig.\ref{fig01}. It should be recognized that a
nonlinear magnetic structure becomes conspicuous only in the vicinity of the
critical field $H_{\text{c}}$.

To see the evolution of the nonlinearity in a more quantitative manner, let us
Fourier decompose the CSL configuration,%
\begin{align}
\cos\varphi_{0}(z)  &  =1+2\left(  \frac{E-K}{\kappa^{2}K}\right) \nonumber\\
&  + \frac{2\pi^{2}}{\kappa^{2}K^{2}}\sum_{n=1}^{\infty}\frac{n\cos\left(
n\pi\bar{z}{/K}\right)  }{\sinh\left(  n\pi K^{\prime}/K\right)
},\label{FourierDecompositionSx}\\
\sin\varphi_{0}(z)  &  =\dfrac{2\pi^{2}}{\kappa^{2}K^{2}}\sum_{n=1}^{\infty
}\dfrac{n\sin\left(  n\pi\bar{z}{/K}\right)  }{\cosh\left(  n\pi K^{\prime
}/K\right)  }, \label{FourierDecompositionSy}%
\end{align}
where $K^{\prime}$ denotes the complete elliptic integral of the first kind
with the complementary elliptic modulus $\kappa^{\prime}\equiv\sqrt
{1-\kappa^{2}}$. We note that $\pi\bar{z}{/K=}2\pi/L_{\text{CSL}}$.

\begin{figure}[ptb]
\begin{center}
\includegraphics[width=70mm,bb=0 0 1000 1200]{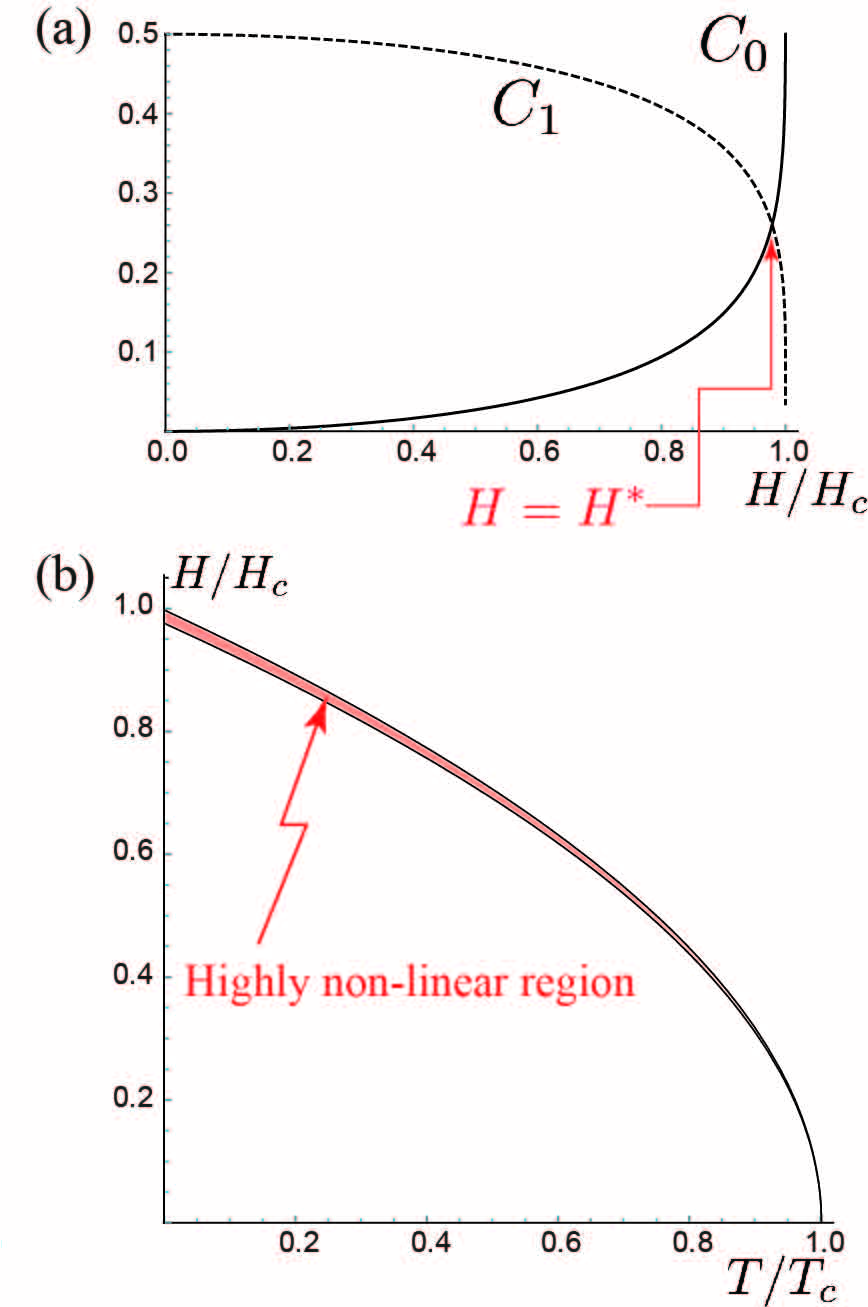}
\end{center}
\caption{(a) Dependence of $C_{0}$ and $C_{1}$ on $H/H_{c}$. They intersect at
$H=H^{\ast}$. (b) Schematic phase diagram where $H_{c}$ and $H^{\ast}$ are
indicated as functions of $T$. The narrow region $H^{\ast}<H<H_{c}$ is
identified with highly non-linear region.}%
\label{fig02}%
\end{figure}

From Eqs. (\ref{FourierDecompositionSx}) and (\ref{FourierDecompositionSy}),
we obtain the Fourier decomposed weights for the ferromagnetic component%
\begin{equation}
C_{0}=\left\{  1+2\left(  \dfrac{E-K}{\kappa^{2}K}\right)  \right\}  ^{2},
\end{equation}
and the spatially modulated components,%
\begin{equation}
C_{n}=\left(  \dfrac{\pi}{\kappa K}\right)  ^{4}\left\{  \dfrac{n^{2}}%
{\sinh^{2}\left(  n\dfrac{{\pi K^{\prime}}}{{K}}\right)  }+\frac{n^{2}}%
{\cosh^{2}\left(  n\dfrac{{\pi K^{\prime}}}{{K}}\right)  }\right\}  .
\end{equation}
The ratio $C_{0}/C_{1}$ is an indicator of nonlinearity in the CSL structure.
The weight $C_{1}$ corresponds to the harmonic modulation of the helix, while
the weight $C_{0}$ indicates an evolution of the ferromagnetic domains. The
dominance of $C_{0}$ over $C_{1}$ means the onset of nonlinearity. In Fig.
\ref{fig02}(a), we show the field dependence of the $C_{0}$\ and $C_{1}$. It
is seen that $C_{1}/C_{0}$ exceeds the unity at $H^{\ast}/H_{c}\simeq0.978$.
This value determines crossover between the linear and the higly nonlinear CSL
regimes.\cite{Laliena2017}

To gain insight into the appearance of the crossover line on the
field-temperature phase diagram, we replace $S$ with a simple mean-field form
having the temperature dependence, $\sqrt{1-T/T_{\text{c}}}$, where $T_{c}$
denotes the transition temperature at zero field. Inserting this form into Eq.
(\ref{critical-field}), we acquire the temperature dependent critical field
$H_{c}(T)=H_{c}(0)\sqrt{1-T/T_{\text{c}}}$. On the other hand, it can be
assumed that the crossover value $H^{\ast}(T)/H_{c}(T)$ is independent on
temperature that brings forth the conceptual phase diagram shown in Fig.
\ref{fig02}(b). We note that the highly nonlinear regime occupies a quite
narrow region bounded by $H^{\ast}(T)<H<H_{c}(T)$.

\subsection{Spectrum of fluctuations}

Below, we thoroughly discuss fluctuations around the soliton lattice ground
state,
\begin{align}
\varphi\left(  z \right)   &  =\varphi_{0}\left(  z\right)  +\delta
\varphi\left(  z \right)  ,\\
\theta\left(  z \right)   &  = \frac{\pi}{2} +\delta\theta\left(  z \right)  .
\end{align}

By expanding the Hamiltonian (\ref{continuumH}) up to the second order with
respect to the $\delta\varphi$ and $\delta\theta$, we obtain $\mathcal{H}%
[\varphi,\theta]=\mathcal{H}[\varphi_{0}]+\delta\mathcal{H}$, where
\begin{equation}
\delta\mathcal{H}=\frac{JS^{2}a_{0}}{2}\int dz\left(  {\delta\varphi}%
\hat{\Lambda}_{{\varphi}}{\delta\varphi+\delta\theta}\hat{\Lambda}_{{\theta}%
}{\delta\theta}\right)  .
\end{equation}
Here, the linear differential operators are given by
\begin{equation}
\hat{\Lambda}_{{\varphi}}=-\left(  \frac{m}{\kappa}\right)  ^{2}\left(
\partial_{\bar{z}}^{2}-2\kappa^{2}\mathrm{sn}^{2}\bar{z}+\kappa^{2}\right)  ,
\label{PhiFluc}%
\end{equation}
and
\begin{equation}
\hat{\Lambda}_{\theta}=\hat{\Lambda}_{\varphi}+\Delta(z) \label{ThetaFluc}%
\end{equation}
with
\begin{equation}
\Delta(z)=-4q_{\text{CSL}}^{2}\mathrm{dn}^{2}\bar{z}+4q_{0}q_{\text{CSL}%
}\mathrm{dn}\bar{z} \label{ThetaGapOriginal}%
\end{equation}
being the energy gap function of the $\theta$-mode originated from the DM
interaction. Here, $q_{\text{CSL}}=2K/L_{\text{CSL}}$ is the wave number of
the CSL structure.

The physical situation of a single kink inside of a ferromagnetic matrix
corresponds to a highly nonlinear regime achieved when $\kappa\rightarrow1$.
In this case, the CSL solution degenerates into
\begin{equation}
\varphi_{0}(\bar{z})\longrightarrow2\pi-2\cos^{-1}\left(  \text{tanh}\bar
{z}\right)  ,
\end{equation}
where $\bar{z}=2z/l_{0}$ with $l_{0}=8/\pi q_{0}$ being the width of the kink
localization (see the inset of Fig.\ref{fig05}).

Then, the operators (\ref{PhiFluc}) and (\ref{ThetaFluc}) take the form
\begin{align}
\hat{\Lambda}_{{\varphi}}  &  \rightarrow\left(  \frac{\pi q_{0}}{4}\right)
^{2}\left(  -\partial_{\bar{z}}^{2}-2\text{\textrm{sech}}^{2}\bar{z}+1\right)
,\\
\hat{\Lambda}_{{\theta}}  &  \rightarrow\left(  \frac{\pi q_{0}}{4}\right)
^{2} \left(  -\partial_{\bar{z}}^{2}-6\text{\textrm{sech}}^{2}\bar{z}
+\frac{16}{\pi}\text{\textrm{sech}}\bar{z}+1\right)  .
\end{align}

\begin{figure}[ptb]
\begin{center}
\includegraphics[width=60mm,bb=0 0 1000 2000]{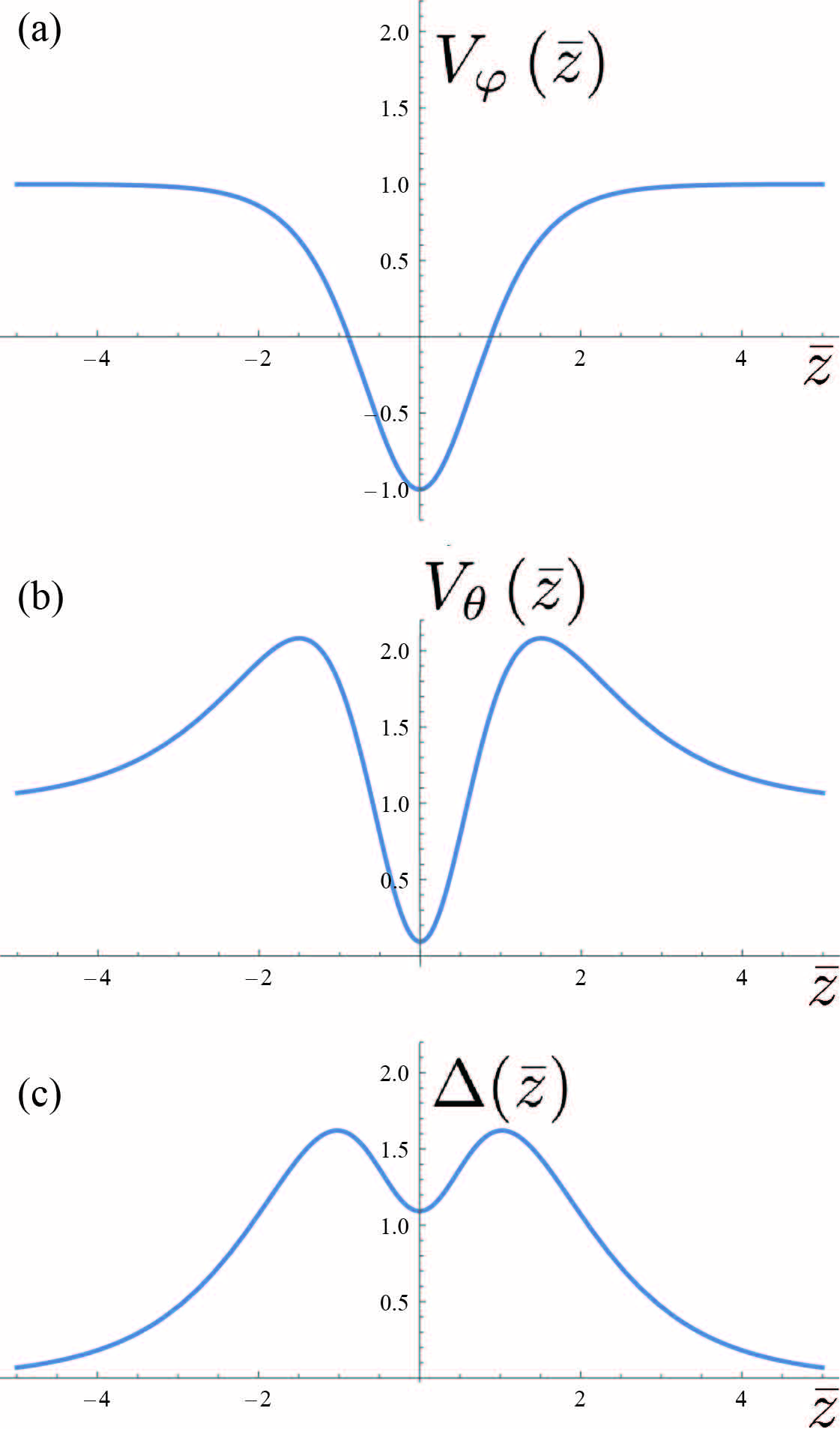}
\end{center}
\caption{Spatial profiles of the potentials for (a) the $\varphi$-fluctuation,
(b) $\theta$- fluctuation, and (c) the gap for the $\theta$- fluctuation, for
the limiting case $\kappa\rightarrow1$.}%
\label{fig03}%
\end{figure}

Figure \ref{fig03} illustrates the potentials for the $\varphi$-fluctuations,
$V_{{\varphi}}\left(  \bar{z}\right)  =-2$\textrm{sech}$^{2}\bar{z}+1$, and
the $\theta$- fluctuations, $V_{{\theta}}\left(  \bar{z}\right)
=-6$\textrm{sech}$^{2}\bar{z}+ \left(  {16}/{\pi}\right)  $\textrm{sech}%
$\bar{z}+1$, where we present additionally the gap function, $\Delta(\bar
{z})=V_{{\theta}}\left(  \bar{z}\right)  -V_{{\varphi}}\left(  \bar{z}\right)
$, at $\kappa=1$.

The Schr\"{o}dinger-type operator $\hat{\Lambda}_{{\varphi}}$ involves the
P\"{o}schl-Teller potential defined by $U_{0}(\bar{z})$ $=-l(l+1)\text{sech}%
^{2}(\bar{z})$ with the particular value $l=1$. For further analysis, the most
important being a presence of the single bound state $\Phi_{0}=2^{-1/2}%
\text{sech}(\bar{z})$ with the eigenvalue $\varepsilon_{0}^{(\varphi)}=0$
(zero mode) \cite{Kishine2010}.

Unfortunately, the operator $\hat{\Lambda}_{\theta}$ does not permit a similar
treatment. However, it may be shown through the WKB formalism (see Appendix A)
that there is a quasi-localized state $u_{0}$ with the energy $\varepsilon
_{0}^{(\theta)}$, $\hat{\Lambda}_{\theta}u_{0}(\bar{z})=\varepsilon
_{0}^{(\theta)}u_{0}(\bar{z})$. By using Eqs.(\ref{WKB_en},\ref{WKB_tp}) we
get $\varepsilon_{0}^{(\theta)}\approx0.0254$ provided $q_{0}=0.16a_{0}^{-1}$.
Numerov algorithm yields $\varepsilon_{0}^{(\theta)}\approx0.0256$. We neglect
henceforth the tunneling process giving a finite width of the state.

\subsection{Lagrangian}

Our target is to obtain equations of motion of the isolated kink in the
ferromagnetic surrounding based on the fluctuation spectra discussed above.

The Lagrangian density includes three terms,
\begin{equation}
\mathcal{L}=\mathcal{L}_{\text{Berry}}+\mathcal{L}_{\text{kink}}%
+\mathcal{L}_{\text{Zeeman}} \label{TLag}%
\end{equation}
such as the kinematic part associated with the Berry phase
\begin{equation}
\mathcal{L}_{\text{Berry}}=\frac{\hslash S}{a_{0}}\int_{-L/2}^{L/2}%
dz(\cos\theta(z)-1)\partial_{t}\varphi(z),
\end{equation}
the part related to the kink energy,
\begin{equation}
\mathcal{L}_{\text{kink}}=\frac{JS^{2}}{2a_{0}}\int_{-L/2}^{L/2}dz\left\{
{\delta\varphi}(z)\hat{\Lambda}_{{\varphi}}{\delta\varphi(z)+\delta\theta
(z)}\hat{\Lambda}_{{\theta}}{\delta\theta}(z)\right\}  ,
\end{equation}
and the Zeeman coupling with the oscillating field, $H_{x}\left(  t\right)
=h_{0}\sin\left(  \Omega t\right)  $, of the strength $h_{0}$ and the
frequency $\Omega$,%
\begin{equation}
\mathcal{L}_{\text{Zeeman}}=-\frac{S}{a_{0}}H_{x}\left(  t\right)  \int
_{-L/2}^{L/2}dz\sin\theta(z)\cos\varphi(z).
\end{equation}
Integration runs over the interval $[-L/2,L/2]$ that has the kink at the center.

In the method of collective coordinates the dynamics is fully described by two
variables, the center-of-mass position $Z(t)$ and the out-of-plane quasi-zero
mode coordinate $\xi_{0}(t)$
\begin{equation}
\label{expanphi0}\varphi(z,t)=\varphi_{0}\left[  z - Z(t)\right]  ,
\end{equation}
\begin{equation}
\label{expantheta0}\theta(z,t)=\pi/2+\xi_{0}(t)u_{0}\left[  z - Z(t)\right]  .
\end{equation}

To obtain equations of motion in the context of the collective coordinates we
expand the Lagrangian (\ref{TLag}) in terms of $\xi$ and $Z$, which are
assumed to be small as long as the ac field is weak.

In this way, we get
\begin{align}
\mathcal{L}  &  \simeq K_{1}\xi_{0}(t)\dot{Z}(t)-\frac{JS^{2}}{2a_{0}%
}\varepsilon_{0}^{(\theta)}\xi_{0}^{2}(t)\nonumber\\
&  +K_{2}H_{x}(t)Z^{2}(t)+K_{3}H_{x}(t)\xi_{0}^{2}(t),
\end{align}
where
\begin{align}
K_{1}  &  =\frac{\hslash S}{a_{0}}\int_{-L/2}^{L/2}d{z}u_{0}\left(  z\right)
\partial_{z}\varphi_{0}\left(  z\right)  ,\\
K_{2}  &  =\frac{S}{2a_{0}}\int_{-L/2}^{L/2}d{z}\left\{  \cos\varphi
_{0}(z)\left[  \varphi_{0}^{^{\prime}}(z)\right]  ^{2}+\sin\varphi
_{0}(z)\varphi_{0}^{^{\prime\prime}}(z)\right\}  ,\\
K_{3}  &  =\dfrac{S}{2a_{0}}\int_{-L/2}^{L/2}d{z}u_{0}^{2}\left(  z\right)
\cos\varphi_{0}(z)
\end{align}
and it is taken into consideration that the sliding coordinate $Z$ corresponds
to the zero mode of $\varphi$-excitations with $\varepsilon_{0}^{(\varphi)}%
=0$. On the other hand, the $\theta$-excitations acquire a finite energy gap
$\varepsilon_{0}^{(\theta)}$.

The equations of motion for the collective coordinates are then given by
\begin{equation}
\label{EOM}%
\begin{array}
[c]{c}%
\dfrac{d\xi_{0}(t)}{dt}=\dfrac{2K_{2}}{K_{1}}h_{0}\sin\left(  \Omega t\right)
Z\left(  t\right)  ,\\
\\
\dfrac{dZ(t)}{dt}= \dfrac{JS^{2} \varepsilon_{0}^{(\theta)}}{K_{1} a_{0}}
\xi_{0}(t)-\dfrac{2K_{3}}{K_{1}}h_{0}\sin\left(  \Omega t\right)  \xi_{0}(t).
\end{array}
\end{equation}

\section{Floquet solution and nonlinearity}

\subsection{Floquet solution}

To find a Floquet solution of these equations of motion it is convenient to
define the integral matrix composed from two linearly independent solutions
\begin{equation}
X = \left(
\begin{array}
[c]{cc}%
\xi^{(1)}_{0} & Z^{(1)}\\
\xi^{(2)}_{0} & Z^{(2)}%
\end{array}
\right)  .
\end{equation}

Then, the system (\ref{EOM}) may be recast into the form
\begin{equation}
\frac{dX}{d\tau}=XP(\tau), \label{IMEQ}%
\end{equation}
where $P(\tau)=P_{0}+h_{0}P_{1}(\tau)$,
\[
P_{0}=\left(
\begin{array}
[c]{cc}%
0 & \rho\\
0 & 0
\end{array}
\right)  ,\quad P_{1}=\left(
\begin{array}
[c]{cc}%
0 & -\beta\\
\alpha & 0
\end{array}
\right)  \sin\tau
\]
with
\begin{align}
\rho &  =4JS^{2}\varepsilon_{0}^{(\theta)}/{(a_{0}\Omega K_{1})},\\
\alpha &  ={2K_{2}}/{(\Omega K_{1})},\\
\beta &  ={2K_{3}}/{(\Omega K_{1})}.
\end{align}
Here, the time $\tau=\Omega t$ is introduced. It is to be noted that $\rho$,
$\alpha$, and $\beta$ are inversely proportional to ${\Omega}$.

A way of constructing Floquet solution of Eq. (\ref{IMEQ}) is explained in the
Appendix B. This method is based on the fact that in the representation of the
integral matrix
\begin{equation}
X(\tau)=\exp\left(  W\tau\right)  N(\tau) \label{FloquetS}%
\end{equation}
the $W$ and $N$ may be expanded as the series with respect to the small
parameter $h_{0}$, that appears in the coefficients of the system
(\ref{IMEQ}). In our analysis the series are limited to third order
\begin{equation}
W\approx W_{0}+h_{0}W_{1}+h_{0}^{2}W_{2}+h_{0}^{3}W_{3}, \label{W}%
\end{equation}%
\begin{equation}
N(\tau)\approx N_{0}(\tau)+h_{0}N_{1}(\tau)+h_{0}^{2}N_{2}(\tau)+h_{0}%
^{3}N_{3}(\tau). \label{N}%
\end{equation}

Following the procedure set out in Appendix C one may find consistently
$W_{0}=P_{0}$ and $N_{0}=I$ [see, Eqs.(\ref{W0},\ref{Z0})]. The explicit forms
of $W_{1}$ and $N_{1}(\tau)$ are originated from Eqs. (\ref{Wk},\ref{Fk}) and
Eq.(\ref{Zk}), respectively,
\begin{equation}
W_{1} = \left(
\begin{array}
[c]{cc}%
-\alpha\rho & 0\\
0 & \alpha\rho
\end{array}
\right)  ,
\end{equation}
\begin{equation}
N_{1} (\tau) = \left(
\begin{array}
[c]{cc}%
\alpha\rho\sin\tau & (\beta- 2\alpha\rho^{2}) (\cos\tau-1)\\
\alpha(1-\cos\tau) & - \alpha\rho\sin\tau
\end{array}
\right)  .
\end{equation}
Reiterating steps of the algorithm for terms of second and third orders (see
Appendix C for details) we obtain the matrix $W$

\begin{widetext}
\begin{equation}
W  = \left(
\begin{array}{cc}
-\frac{5}{12} \left(7 \alpha ^3 \rho ^3-4 \alpha ^2 \beta  \rho \right) h^3_0 - \alpha
\rho  h_0  & \frac{1}{4} \left(5 \alpha ^2 \rho ^3-6 \alpha  \beta  \rho \right)
h^2_0+\rho  \\
-\frac{3}{2} \alpha ^2 h^2_0 \rho  & \frac{5}{12} \left(7 \alpha ^3 \rho ^3-4 \alpha ^2
\beta  \rho \right) h^3_0+\alpha  \rho  h_0  \\
\end{array}
\right).
\end{equation}
\end{widetext}
It has purely imaginary characteristic numbers that yields
\begin{equation}
\exp\left(  W \tau\right)  = \cos(\lambda\tau) I + \frac{1}{\lambda}
\sin(\lambda\tau) W,
\end{equation}
with
\[
\lambda= \frac{1}{\sqrt{2}} \left[  \alpha\rho h_{0} - \frac{1}{24} h^{3}_{0}
\left(  95 \alpha^{3} \rho^{3}-26 \alpha^{2} \beta\rho\right)  \right]  ,
\]
and the integral matrix $X$ turns out to be oscillatory as a result.

By carrying out direct calculation of Eq. (\ref{FloquetS}) and ignoring a
frequency shift to the value $\lambda\ll1$, we get eventually the first pair
of the Floquet solution \begin{widetext}
$$
\xi^{(1)}_0 (\tau)  = 1+ \frac{11}{8} \alpha ^2 \rho ^2 h^2_0 -\frac{3}{4} \alpha  \beta  h^2_0
+ \alpha  h^2_0 \left(\beta -2 \alpha  \rho ^2\right)  \cos \tau
+ \frac{1}{8} \alpha  \rho  h_0  \left(17 \alpha ^2 \rho ^2 h^2_0 - 8 \alpha  \beta
h ^2_0 +8\right) \sin \tau
$$
\begin{equation}
+ \frac{1}{8} \alpha  h^2_0 \left(5 \alpha  \rho ^2-2 \beta \right) \cos 2\tau
+ \frac{1}{4} \alpha ^2 \rho  h^3_0 \left(2 \alpha  \rho ^2-\beta \right) \sin 2\tau
-\frac{1}{72} \alpha ^2 \rho  h^3_0  \left(5 \alpha  \rho ^2+4 \beta \right) \sin 3\tau,
\end{equation}
$$
Z^{(1)} (\tau) = -\beta  h_0 +2 \alpha  \rho ^2 h_0 +\frac{5}{12} \alpha  \beta ^2 h^3_0  -\frac{719}{216} \alpha ^2 \beta  \rho ^2 h^3_0 + \frac{571}{108} \alpha ^3 \rho ^4 h^3_0
$$
$$
-\frac{1}{16} h_0  \left(81 \alpha ^3 \rho ^4 h^2_0 -57 \alpha ^2 \beta  \rho ^2
h^2_0+10 \alpha  \beta ^2 h^2_0+32 \alpha  \rho ^2-16 \beta \right)  \cos \tau
+ \alpha  \rho  h^2_0 \left(\beta -2 \alpha  \rho ^2\right) \sin \tau
$$
\begin{equation}
-\frac{1}{8} \alpha  h^3_0 \left(2 \alpha ^2 \rho ^4+3 \alpha  \beta  \rho ^2-2 \beta
^2\right) \cos 2\tau
+ \frac{1}{8} \alpha  \rho  h^2_0  \left(3 \alpha  \rho ^2+2 \beta \right) \sin 2\tau
+ \frac{1}{432} \alpha  h^3_0 \left(11 \alpha ^2 \rho ^4+61 \alpha  \beta  \rho ^2-18 \beta
^2\right) \cos 3\tau.
\end{equation}
Similarly, we find the second pair
$$
\xi^{(2)}_0 (\tau)  = \alpha   h_0 -\frac{5}{12} \alpha ^2 \beta  h^3_0 + \frac{49}{24} \alpha ^3 \rho ^2   h^3_0
-\frac{1}{16} \alpha  h_0  \left(43 \alpha ^2 \rho ^2  h^2_0 -10 \alpha  \beta
h^2_0 +16\right) \cos \tau
$$
\begin{equation}  \label{XI0sol}
+ \alpha ^2 \rho  h^2_0 \sin \tau + \frac{1}{8} \alpha ^2 h^3_0 \left(5 \alpha  \rho ^2-2 \beta \right) \cos 2 \tau
+ \frac{1}{4} \alpha ^2 \rho  h^2_0 \sin 2 \tau
+ \frac{1}{48} \alpha ^2 h^3_0 \left(\alpha  \rho ^2+2 \beta \right) \cos 3 \tau,
\end{equation}
$$
Z^{(2)}   (\tau) =  1 -\frac{3}{4} \alpha  \beta  h^2_0 + \frac{17}{8} \alpha ^2 \rho ^2 h^2_0
+ \alpha  h^2_0 \left(\beta -2 \alpha  \rho ^2\right) \cos \tau
+ \left( -\alpha  \rho    h_0 +\alpha ^2 \beta  \rho  h^3_0 -\frac{59}{16} \alpha ^3 \rho ^3 h^3_0 \right) \sin \tau
$$
\begin{equation}  \label{Z0sol}
-\frac{1}{8} \alpha  h^2_0 \left(\alpha  \rho ^2+2 \beta \right) \cos 2 \tau
+ \frac{1}{8} \alpha ^2 \rho  h^3_0 \left(3 \alpha  \rho ^2+2 \beta \right) \sin 2 \tau
+ \frac{1}{144} \alpha ^2 \rho  h^3_0 \left(\alpha  \rho ^2+8 \beta \right) \sin 3 \tau,
\end{equation}
\end{widetext}
which is the physical solution consistent with the initial condition $\xi
_{0}=0$, $Z=1$. Bearing in mind that if $X(\tau)$ is the normalized integral
matrix at the point $\tau=0$, i.e. $X(0)=I$, then every other integral matrix
$\tilde{X}(\tau)$ can be expressed in the form $\tilde{X}(\tau) = A X(\tau)$,
where $A$ is a constant matrix. Therefore, transition to an arbitrary kink
position $Z_{0}$ is achieved by the matrix $A=\text{diag} \left\{  1, Z_{0}
\right\}  $.

\subsection{Magnetization}

The resultant magnetization is originated from dependence on the collective
coordinates
\begin{gather}
M_{x}(t)/M_{0}=\int_{-L/2}^{L/2}\dfrac{dz}{a_{0}}\sin\theta(z,t)\cos
\varphi(z,t)\nonumber\\
=\int_{-L/2}^{L/2}\dfrac{dz}{a_{0}}\cos\left[  \xi_{0}(t)u_{0}\left(
z-Z(t)\right)  \right]  \cos\left[  \varphi_{0}\left(  z-Z(t)\right)  \right]
,
\end{gather}
where $M_{0}$ is the magnetization amplitude.

Performing expansion in powers of $\xi^{m}Z^{n}$ right up to third order
($m+n\leq3$), we obtain the fluctuation part,
\begin{equation}
\delta M_{x}/M_{0}=-\frac{a}{2}Z^{2}-\frac{b}{2}\xi_{0}^{2}+\frac{c}{6}%
Z^{3}+\frac{f}{2}Z\xi_{0}^{2}, \label{MagExpan}%
\end{equation}
where
\begin{align}
a  &  =\int_{-L/2}^{L/2}\dfrac{dz}{a_{0}}\left\{  \varphi_{0}^{\prime\prime
}(z)\sin\varphi_{0}(z)+\varphi_{0}^{\prime2}\cos\varphi_{0}(z)\right\}  ,\\
b  &  =\int_{-L/2}^{L/2}\dfrac{dz}{a_{0}}\,u_{0}(z)^{2}\cos\varphi_{0}(z),\\
c  &  =\int_{-L/2}^{L/2}\dfrac{dz}{a_{0}}\left\{
\begin{array}
[c]{c}%
\varphi_{0}^{(3)}(z)\sin\varphi_{0}(z)-\varphi_{0}^{\prime3}\sin\varphi
_{0}(z)\\
+3\varphi_{0}^{\prime}(z)\varphi_{0}^{\prime\prime}(z)\cos\varphi_{0}(z)
\end{array}
\right\}  ,\\
f  &  =\int_{-L/2}^{L/2}\dfrac{dz}{a_{0}}\left\{
\begin{array}
[c]{c}%
2u_{0}(z)u_{0}^{\prime}(z)\cos\varphi_{0}(z)\\
-u_{0}(z)^{2}\varphi_{0}^{\prime}(z)\sin\varphi_{0}(z)
\end{array}
\right\}  .
\end{align}
It follows on parity grounds that $c=0$ and $f=0$, note aslo that $a=2K_{2}/S$
and $b=2K_{3}/S$.

By plugging the solutions (\ref{XI0sol},\ref{Z0sol}) obtained earlier in
(\ref{MagExpan}) and neglecting higher-order terms the magnetization as
function on time may be presented as follows
\begin{equation}
\delta M_{x}(t)/M_{0}=\sum_{n=1}^{3}M_{n0}\sin\left(  n\Omega t+\delta
_{n}\right)  .
\end{equation}
Here, the $n$th-harmonic components, $n=1\ldots3$, are given by
\begin{align}
M_{10}  &  =|a\alpha\rho|h_{0},\label{M10}\\
M_{20}  &  =\frac{1}{8}|\alpha\left(  2a\beta-2b\alpha+3a\alpha\rho
^{2}\right)  |h_{0}^{2},\label{M20}\\
M_{30}  &  =\frac{1}{72}\alpha^{2}|\rho\left(  13a\beta-9b\alpha+5a\alpha
\rho^{2}\right)  |h_{0}^{3}. \label{M30}%
\end{align}
We note that $M_{10},$ $M_{20}$ are proportional to $\Omega^{-2}$, while
$M_{30}$ to $\Omega^{-4}$, since $\rho$, $\alpha$, and $\beta$ are inversely
proportional to ${\Omega}$.

The expressions for
\begin{align}
\delta_{1} = \tan^{-1} \left[  \frac{ h_{0}(-a \beta+b \alpha+ 2 a \alpha
\rho^{2})}{a \rho} \right]  ,
\end{align}
\begin{align}
\delta_{2} = \tan^{-1} \left[  \frac{(2a\beta- 2b \alpha+ 3a \alpha\rho^{2}%
)}{\alpha\rho h_{0}(2a \beta+ 2b \alpha-11 a \alpha\rho^{2})} \right]  ,
\end{align}
and $\delta_{3}=0$ determine the phase delay of each $M_{n0}$ against the ac-field.

Given the characteristic length of the \textquotedblleft kink
localization\textquotedblright, $l_{0}$, we take the size of the system as
\begin{equation}
L/2=Nl_{0}=8N/\pi q_{0}=8JNa_{0}/\pi D\propto Na_{0},
\end{equation}
where $N$ measures how large the whole system size is as compared with the
kink width. Larger $N$ corresponds to more sparse distribution of the kinks.
Then, $K_{1}$, $K_{2}$ and $K_{3}$ obey the following scaling laws with
respect to $N$,
\begin{align}
K_{1}  &  \sim\frac{\hbar S}{a_{0}}\frac{1}{\sqrt{L}}\int_{-L/2}^{L/2}%
\partial_{z}\varphi_{0}\left(  z\right)  dz\nonumber\\
&  =\frac{\hbar S}{a_{0}}\frac{2\pi}{\sqrt{L}}\sim\hbar Sa_{0}^{-3/2}%
N^{-1/2},\\
K_{2}  &  =\frac{S}{2a_{0}}\int_{-L/2}^{L/2}d{z}\frac{{d}}{dz}\left\{
\varphi_{0}^{^{\prime}}(z)\sin\varphi_{0}(z)\right\} \nonumber\\
&  =\frac{S}{a_{0}}\left(  \frac{\pi q_{0}}{2}\right)  \text{sech}2N\sim
Sa_{0}^{-2}\text{sech}2N,\\
K_{3}  &  \sim\dfrac{S}{2a_{0}}\int_{-L/2}^{L/2}d{z}u_{0}^{2}\left(  z\right)
\cos\varphi_{0}(z)\sim Sa_{0}^{-1},
\end{align}
where we used $u_{0}\sim1/\sqrt{L}$ (see Ref. \cite{Kishine2010}) and
$\sin\varphi_{0}(L/2)\sim1$. We note the parameters $a$ and $b$\ has the same
dependence on $N$ with $K_{2}$ and $K_{3}$, respectively.

\begin{figure}[ptb]
\begin{center}
\includegraphics[width=80mm,bb=0 0 1000 1200]{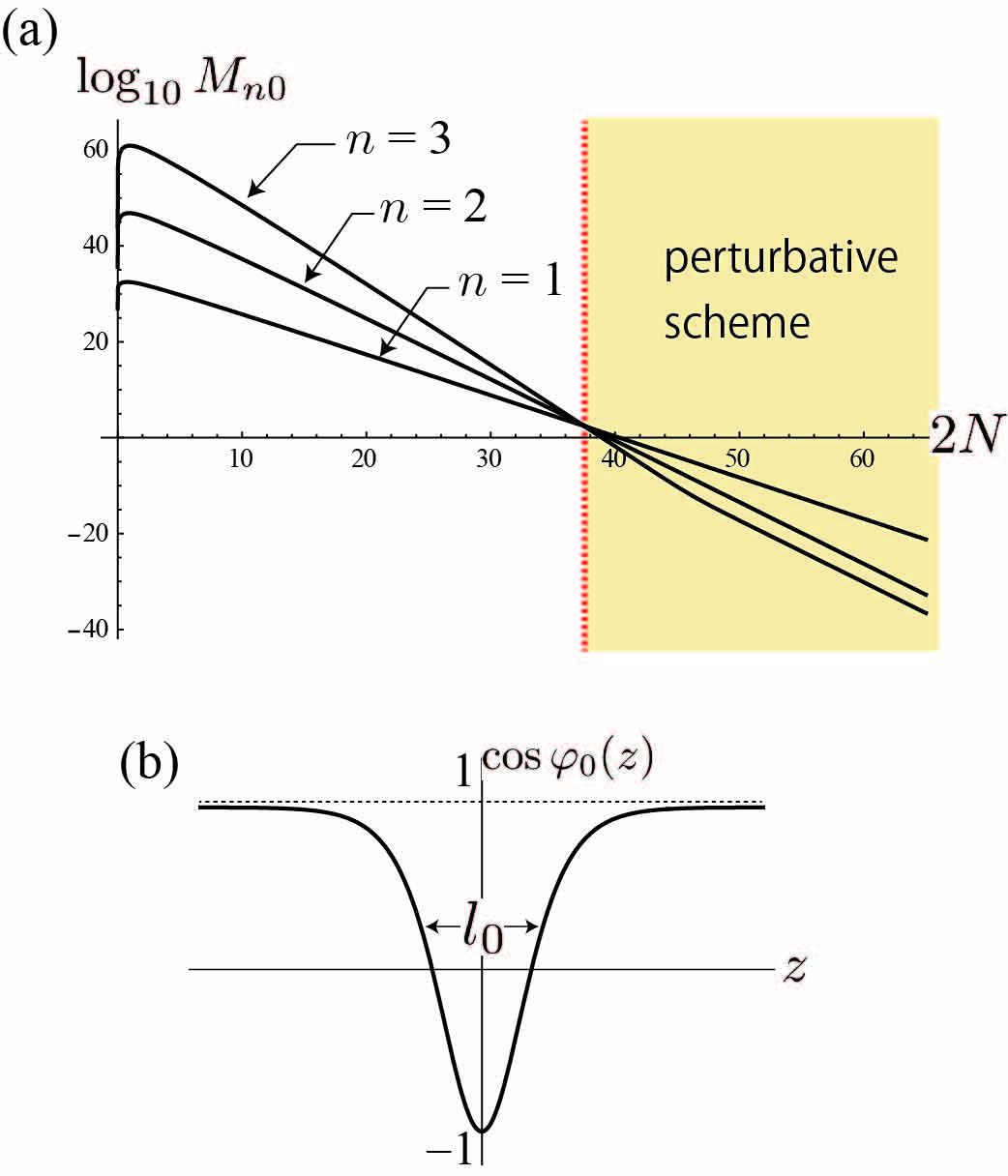}
\end{center}
\caption{(a) Scale behavior of the amplitude $M_{n0}$ ($n=1,2,3$) with a
growth of $L/a_{0}\sim2N$. The dotted vertical line marks an onset of
applicability of the perturbation analysis. (b) The spatial profile of a
single kink with $l_{0}=8/\pi q_{0}$ being the width of the kink
localization.}%
\label{fig05}%
\end{figure}

By using the values for CrNb$_{3}$S$_{6}$ as $S=3/2$, $J=18$ K (or
$2.484\times10^{-22}$ J), $a_{0}=1.212\times10^{-9}$ m. As for the ac magnetic
field, we follow Ref. \cite{Tsuruta2016} and take $h_{0}=5$ Oe (then
$g\mu_{\text{B}}h_{0}=9.274\times10^{-27}$ J) and $\nu=2\pi\Omega=10$ Hz (then
$h\nu=6.626\times10^{-33}$ J). These vales lead to order-of-magnitude
estimate,%
\begin{align}
\alpha h_{0}  &  ={2K_{2}}h_{0}/{(\Omega K_{1})}\sim10^{10}\sqrt{N}%
\text{sech}(2N),\\
\beta h_{0}  &  ={2K_{3}}h_{0}/{(\Omega K_{1})}\sim10\sqrt{N},\\
\rho &  =4JS^{2}\varepsilon_{0}^{(\theta)}/{(a_{0}\Omega K_{1})}\sim
10^{4}\sqrt{N}.
\end{align}
Plugging them into Eqs. (\ref{M10}-\ref{M30}), we see that the harmonic
components obey the scaling laws with respect to $N$ as
\begin{align}
M_{10}  &  \sim10^{33}N\text{sech}^{2}(2N),\\
M_{20}  &  \sim10^{48}N^{2}\text{sech}^{3}(2N),\\
M_{30}  &  \sim10^{62}N^{3}\text{sech}^{4}(2N).
\end{align}
We show the $N$-dependence of $\log_{10}M_{n0}$ in Fig. \ref{fig05}(a). By
choosing the characteristic length per the kink, $L/a_{0}\sim2N$, as $2N=41$,
we obtain the proportion close to that observed in the experiment $M_{10}%
\sim0.2728$, $M_{20}\sim0.0131$ and $M_{30}\sim0.0003$. The corresponding
phase shifts are $\delta_{1}\sim0.251$ and $\delta_{2}\sim-1.132$.

It is seen that as functions of $N$, the amplitudes of higher harmic
contributions, $M_{20}$ and\ $M_{30}$ dominate $M_{10}$ for smaller
$2N\lesssim40$, which indicates that the perturbative scheme [expansions given
in Eqs. (\ref{W}) and (\ref{N})] breaks down. Because larger $2N$ means lower
kink density, our scheme works for the regime of smaller kink density
specified by $2N\gtrsim40$. This condition is consistent with the situation
where the highly non-linear regime is near the boundary of the nucleation
transition [see Fig. \ref{fig02}(b)].

\section{Conclusion and discussions}

The accumulated data on nonlinear response in CrNb$_{3}$S${}_{6}$ provide
exciting challenges that need to be treated theoretically. This task is
closely linked with a general issue of emergence of slow dynamics from high
energy processes \cite{Fukuyama2017}. In our study, we explain the origin of
this phenomenon in the regime of highly nonlinear soliton lattice by internal
deformations of separate $2\pi$-kinks driven by an external ac magnetic field.
We demonstrate that the emergence of higher-order harmonics takes place in a
narrow range of dc fields when there is an optimal distance between the kinks.
At lower distances that corresponds to high density of the kinks our analysis
based on a perturbative scheme becomes irrelevant. For larger lengths, i.e.
small kink density, contribution of the higher-order harmonics is negligible
and we have linear magnetic response in the vicinity of the phase transition
into the state of forced ferromagnetism.

Note that temperature effects related both with the nonlinear response and
behavior of the chiral soliton lattice as a
whole\cite{Krumhansl1975,Gupta1976,Shinozaki2016} remain beyond our treatment.
The presented theory nevertheless may be exploited in addressing of
second-order phase transitions of the nucleation type.

\begin{acknowledgments}
The authors would like to express special thanks to Profs. Masaki Mito,
Manh-Huong Phan, David Mandrus and Hidetoshi Fukuyama for very informative
discussions during various stages. The authors also thank Victor Laliena,
Javier Campo, and Yusuke Kato for fruitful discussions. This work was
supported by a Grant-in-Aid for Scientific Research (B) (No. 17H02923) from
the MEXT of the Japanese Government. A.S.O. acknowledges funding by the
Foundation for the Advancement of Theoretical Physics and Mathematics BASIS
Grant No. 17-11-107, and by Act 211 Government of the Russian Federation,
contract No. 02.A03.21.0006. A.S.O. thanks also the Ministry of Education and
Science of the Russian Federation, Project No. 3.2916.2017/4.6.
\end{acknowledgments}

\appendix

\section{WKB method for $V_{\theta}$}

\begin{figure}[ptb]
\begin{center}
\includegraphics[width=80mm,bb=10 0 974 529]{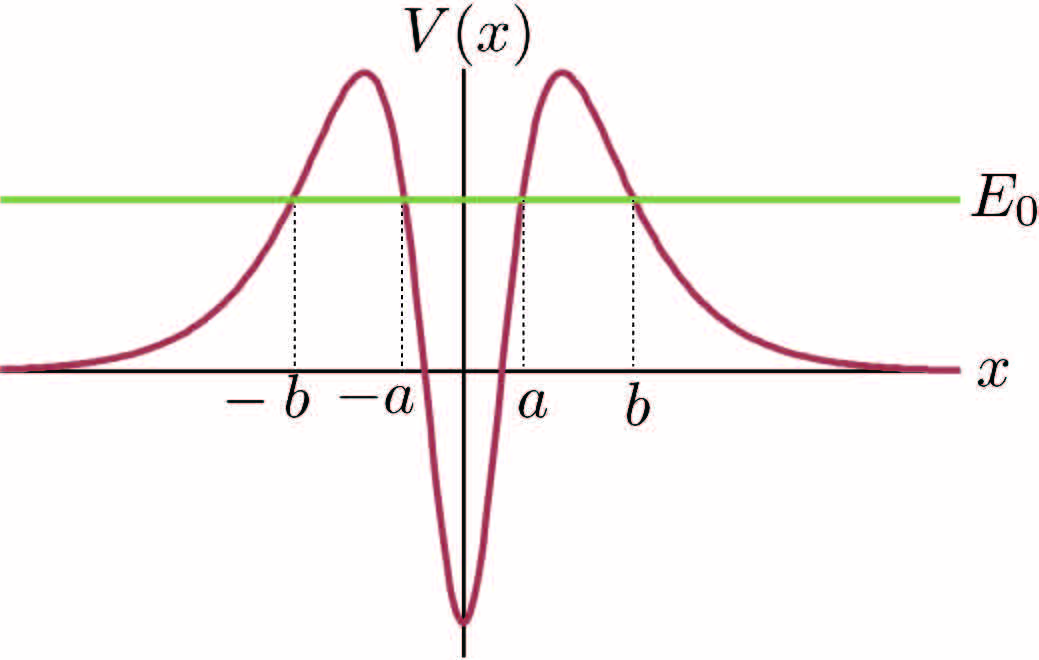}
\end{center}
\caption{The inverse double well potential for the WKB analysis.}%
\label{fig06}%
\end{figure}

Below, we find the quasi-stationary levels of a particle in the symmetrical
potential shown in Fig. \ref{fig06} following a general scheme outlined in
Ref. \cite{Haar1964}.

In the region $x<-b$ we have a wave which goes to $-\infty$,
\begin{equation}
\psi=\frac{c}{\sqrt{p}}\exp\left(  \frac{i}{\hbar}\int_{x}^{-b}pdx\right)  ,
\end{equation}
where $c$ is a constant and $p$\ is a momentum of the particle. In the region
$-b<x<-a$ we obtain
\begin{equation}
\psi=\frac{c}{2\sqrt{|p|}}\exp\left(  \frac{i\pi}{4}-\frac{1}{\hbar}\int
_{-b}^{-a}|p|dx+\frac{1}{\hbar}\int_{x}^{-a}|p|dx\right)
\end{equation}%
\[
+\frac{c}{\sqrt{|p|}}\exp\left(  -\frac{i\pi}{4}+\frac{1}{\hbar}\int_{-b}%
^{-a}|p|dx-\frac{1}{\hbar}\int_{x}^{-a}|p|dx\right)  .
\]
In the region $-a<x<a$ we get
\begin{equation}
\psi=\frac{c}{\sqrt{p}}\left[  \frac{1}{4}\exp\left(  \frac{i\pi}{2}-\frac
{1}{\hbar}\int_{-b}^{-a}|p|dx\right)  \right.
\end{equation}%
\[
\left.  +\exp\left(  -\frac{i\pi}{2}+\frac{1}{\hbar}\int_{-b}^{-a}%
|p|dx\right)  \right]
\]%
\[
\times\exp\left(  \frac{i}{\hbar}\int_{-a}^{a}pdx-\frac{i}{\hbar}\int_{x}%
^{a}pdx\right)
\]%
\[
\frac{c}{\sqrt{p}}\left[  \frac{1}{4}\exp\left(  -\frac{1}{\hbar}\int
_{-b}^{-a}|p|dx\right)  +\exp\left(  \frac{1}{\hbar}\int_{-b}^{-a}%
|p|dx\right)  \right]
\]%
\[
\times\exp\left(  -\frac{i}{\hbar}\int_{-a}^{a}pdx+\frac{i}{\hbar}\int_{x}%
^{a}pdx\right)  .
\]
In the region $a<x<b$
\begin{equation}
\psi=\frac{c}{\sqrt{|p|}}\exp\left(  -\frac{1}{\hbar}\int_{a}^{b}%
|p|dx+\frac{1}{\hbar}\int_{x}^{b}|p|dx\right)
\end{equation}%
\[
\times\left[  \sin\left(  \frac{1}{\hbar}\int_{-a}^{a}pdx\right)  \exp\left(
-\frac{i\pi}{4}+\frac{1}{\hbar}\int_{-b}^{-a}|p|dx\right)  \right.
\]%
\[
\left.  +\frac{1}{4}\cos\left(  \frac{1}{\hbar}\int_{-a}^{a}pdx\right)
\exp\left(  \frac{i\pi}{4}-\frac{1}{\hbar}\int_{-b}^{-a}|p|dx\right)  \right]
\]%
\[
+\frac{c}{\sqrt{|p|}}\exp\left(  \frac{1}{\hbar}\int_{a}^{b}|p|dx-\frac
{1}{\hbar}\int_{x}^{b}|p|dx\right)
\]%
\[
\times\left[  -\frac{1}{2}\sin\left(  \frac{1}{\hbar}\int_{-a}^{a}pdx\right)
\exp\left(  \frac{i\pi}{4}-\frac{1}{\hbar}\int_{-b}^{-a}|p|dx\right)  \right.
\]%
\[
\left.  +2\cos\left(  \frac{1}{\hbar}\int_{-a}^{a}pdx\right)  \exp\left(
-\frac{i\pi}{4}+\frac{1}{\hbar}\int_{-b}^{-a}|p|dx\right)  \right]  .
\]
In the region $x>b$ we find
\begin{equation}
\psi=\frac{c}{\sqrt{p}}\exp\left(  \frac{i}{\hbar}\int_{b}^{x}pdx\right)
\end{equation}%
\[
\times\left[  \frac{1}{8}\exp\left(  -\frac{2}{\hbar}\int_{a}^{b}%
|p|dx+\frac{i\pi}{2}\right)  \cos\left(  \frac{1}{\hbar}\int_{-a}%
^{a}pdx\right)  \right.
\]%
\[
\left.  +2\exp\left(  \frac{2}{\hbar}\int_{a}^{b}|p|dx-\frac{i\pi}{2}\right)
\cos\left(  \frac{1}{\hbar}\int_{-a}^{a}pdx\right)  \right]
\]%
\[
+\frac{c}{\sqrt{p}}\exp\left(  -\frac{i}{\hbar}\int_{b}^{x}pdx\right)
\]%
\[
\times\left[  \frac{1}{8}\exp\left(  -\frac{2}{\hbar}\int_{a}^{b}|p|dx\right)
\cos\left(  \frac{1}{\hbar}\int_{-a}^{a}pdx\right)  \right.
\]%
\[
\left.  +2\exp\left(  \frac{2}{\hbar}\int_{a}^{b}|p|dx\right)  \cos\left(
\frac{1}{\hbar}\int_{-a}^{a}pdx\right)  -i\sin\left(  \frac{1}{\hbar}\int
_{-a}^{a}pdx\right)  \right]  .
\]

An absence of a wave coming from $+\infty$ yields
\begin{equation}
\text{cot}\left(  \frac{1}{\hbar}\int_{-a}^{a}pdx\right)
\end{equation}%
\[
=i\left[  \frac{1}{8}\exp\left(  -\frac{2}{\hbar}\int_{a}^{b}|p|dx\right)
+2\exp\left(  \frac{2}{\hbar}\int_{a}^{b}|p|dx\right)  \right]  ^{-1}.
\]
Provided $\exp\left(  -\frac{2}{\hbar}\int_{a}^{b}|p|dx\right)  \ll1$, we
obtain
\begin{equation}
\frac{1}{\hbar}\int_{-a}^{a}pdx=\pi\left(  n+\frac{1}{2}\right)  -\frac{i}%
{2}\exp\left(  -\frac{2}{\hbar}\int_{a}^{b}|p|dx\right)  ,
\end{equation}
where $n$ is a non-negative integer.

The quasi-stationary levels $E_{n}^{(0)}$ and their width $\Gamma_{n}$ are
given by
\begin{equation}
\frac{1}{\hbar}\int_{-a}^{a}\sqrt{2m\left(  E_{n}^{(0)}-V(x)\right)  }%
dx=\pi\left(  n+\frac{1}{2}\right)  , \label{En}%
\end{equation}
where $V(x)$\ is the potential, $m$ is the mass of the particle, and
\begin{equation}
\Gamma_{n}=\frac{\hbar\omega}{2\pi}\exp\left(  -\frac{2}{\hbar}\int_{a}%
^{b}|p|dx\right)  .
\end{equation}
Here, $\omega$ is the angular frequency of the classical motion in a separate
well,
\[
\omega=\pi\left(  m\int_{-a}^{a}\left[  2m\left(  E_{n}^{(0)}-V(x)\right)
\right]  ^{-\frac{1}{2}}dx\right)  ^{-1}.
\]

To find $\varepsilon^{(\theta)}_{0}$ we consider the Schroedinger-like
equation
\begin{equation}
\label{SchEq}\left\{  -\frac12 \frac{d^{2}}{dz^{2}} - \frac{3}{\cosh^{2} z} +
\frac{8}{\pi\cosh z} \right\}  u_{0}(z) = E_{0} u_{0}(z)
\end{equation}
with $E_{0} = \left[  8 \varepsilon^{(\theta)}_{0} /(\pi q_{0} a_{0})^{2} -
1/2 \right]  $.

The energy of the quasi-localized state may found from (\ref{En}) at $n=0$,
\begin{equation}
\int_{0}^{a}\sqrt{E_{0}+\frac{3}{\cosh^{2}z}-\frac{8}{\pi\cosh z}}dz=\frac
{\pi}{4\sqrt{2}}, \label{WKB_en}%
\end{equation}
where the upper limit of integration is related with the $E_{0}$
\begin{equation}
E_{0}=-\frac{3}{\cosh^{2}a}+\frac{8}{\pi\cosh a}, \label{WKB_tp}%
\end{equation}
that results in $E_{0}=0.30581$. \begin{figure}[ptb]
\begin{center}
\includegraphics[width=80mm,bb=0 0 1000 600]{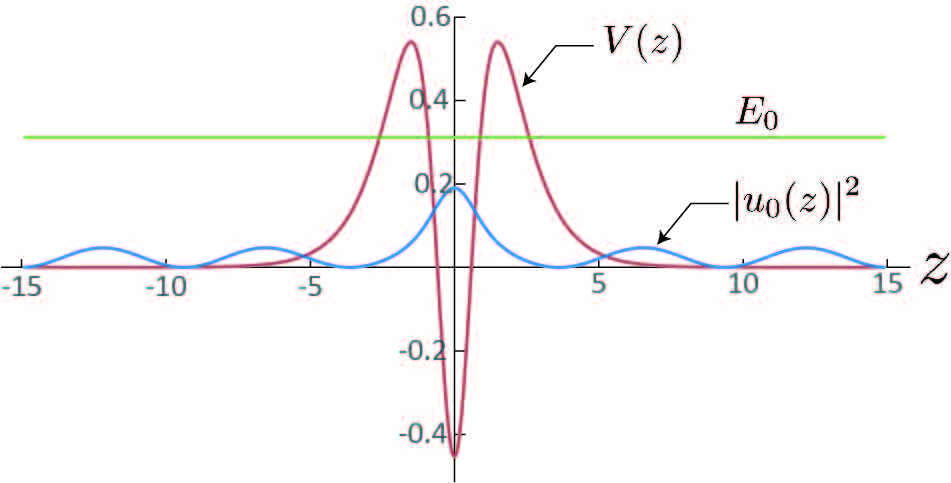}
\end{center}
\caption{Spatial profiles of the potential $V(z)$ (red) for the
quasi-localized state, the probability density $\left\vert u_{0}(z)\right\vert
^{2}$ (blue) and the corresponding energy $E_{0}$ (green) obtained by the
Numerov's algorithm.}%
\label{fig04}%
\end{figure}In Fig. \ref{fig04} the numerical solution via the Numerov's
algorithm is shown\cite{Landau2008}. The corresponding $E_{0}$ value is
$0.31121$.

\section{Erugin's method}

We consider a system of the form
\begin{equation}
\frac{dX}{dt}=X\sum_{k=0}^{\infty}P_{k}(t)\varepsilon^{k}, \label{FSystem}%
\end{equation}
where $P_{k}(t)$ are $n$-th order matrices that are continuous and periodic
with period $2\pi$, $\varepsilon$\ is a small parameter.

The integral matrix of Eq.(\ref{FSystem}) normalized at the point $t=0$ can be
expressed as a series
\begin{equation}
X(t) = \sum_{k=0}^{\infty} X_{k} (t) \varepsilon^{k},
\end{equation}
with $X_{0}(0)=1$, $X_{k}(0)=0$ at $k \geq1$.

It can be shown (see Ref. \cite{Erugin}) that the integral matrix, giving
Floquet solution, can be represented in the form
\begin{equation}
X(t,\varepsilon) = \exp\left(  W(\epsilon) t \right)  N(t,\varepsilon).
\end{equation}
where $W(\epsilon)$ is the real constant matrix and $N(t,\varepsilon)$ is
periodic with the period $2\pi$.

According to the general theory we have these quantities in the form of the
series in powers of $\epsilon$,
\begin{equation}
W(\varepsilon) = \sum_{k=0}^{\infty} W_{k} \varepsilon^{k}, \qquad
N(t,\varepsilon) = \sum_{k=0}^{\infty} N_{k}(t) \varepsilon^{k}.
\end{equation}

The recipe for finding of the $W_{k}$ and $N_{k}(t)$ may be explained as
follows. Firstly, we define
\begin{equation}
\label{W0}W_{0} = \frac{1}{2\pi} \ln\left[  \exp\left(  2\pi P_{0} \right)
\right]  ,
\end{equation}
and
\begin{equation}
\label{Z0}N_{0}(t) = e^{-W_{0} t} \exp\left(  P_{0} t \right)  .
\end{equation}

We can now calculate the periodic matrix $F_{k}(t)$ with period $2\pi$
\begin{equation}
\label{Fk}F_{k}(t) = \sum_{\nu=1}^{k} N_{k-\nu}(t) P_{\nu}(t) - \sum_{\nu
=1}^{k-1} W_{k-\nu} N_{\nu}(t).
\end{equation}

Then, $W_{k}$ may be found from
\[
\int^{2\pi}_{0} \exp{\left(  P_{0} t \right)  } F_{k} N^{-1}_{0}(t)
\exp{\left(  - P_{0} t \right)  } dt
\]
\begin{equation}
\label{Wk}= \int^{2\pi}_{0} \exp{\left(  P_{0} t \right)  } W_{k} \exp{\left(
- P_{0} t \right)  } dt.
\end{equation}
After all, we obtain
\[
N_{k}(t) = \exp{\left(  - P_{0} t \right)  } \left[  \int^{t}_{0} \exp{\left(
P_{0} t^{\prime}\right)  } \left(  F_{k}(t^{\prime-1}_{0}(t^{\prime}) \right.
\right.
\]
\begin{equation}
\label{Zk}\left.  \left.  -W_{k} \right)  \exp{\left(  - P_{0} t^{\prime
}\right)  } dt^{\prime}\right]  \exp{\left(  P_{0} t \right)  } N_{0}(t).
\end{equation}

\section{Matrices $W_{k}$ and $N_{k}$ for $k=2,3$}

Below we result explicitly the matrices $W_{k}$ and $Z_{k}(\tau)$, $k=2,3$,
necessary to build a Floquet solution.

By using (\ref{Fk}-\ref{Zk}) one may find \begin{widetext}
\begin{equation}
W_2  = \left(
\begin{array}{cc}
0  & \frac14 \alpha \rho (-6\beta + 5 \alpha \rho^2)  \\
-\frac32 \alpha^2 \rho &  0
\end{array}
\right),  \quad
W_3  = \left(
\begin{array}{cc}
-\frac{5}{12} \left(7 \alpha ^3 \rho ^3-4 \alpha^2 \beta  \rho \right)   & 0  \\
0 & \frac{5}{12} \left(7 \alpha ^3 \rho ^3-4 \alpha ^2
\beta  \rho \right)
\end{array}
\right),
\end{equation}
\begin{equation}
N_2(\tau)  = \left(
\begin{array}{cc}
N^{(2)}_{11}  &  N^{(2)}_{12}  \\
N^{(2)}_{21}  & N^{(2)}_{22}
\end{array}
\right), \quad
N_3(\tau)  = \left(
\begin{array}{cc}
N^{(3)}_{11}  &  N^{(3)}_{12}  \\
N^{(3)}_{21}  & N^{(3)}_{22}
\end{array}
\right)
\end{equation}
with
$$
N^{(2)}_{11} = \frac{1}{8} \left[11 \alpha ^2 \rho ^2-6 \alpha  \beta -16 \alpha ^2 \rho ^2 \cos \tau  +8 \alpha  \beta  \cos \tau +5 \alpha ^2 \rho ^2 \cos 2 \tau -2 \alpha
\beta  \cos 2 \tau \right],
$$
$$
N^{(2)}_{12} =   \frac{1}{8} \left[-16 \alpha ^2 \rho ^3 \sin \tau +8 \alpha  \beta  \rho  \sin \tau +3 \alpha ^2 \rho ^3 \sin 2 \tau +2 \alpha  \beta  \rho  \sin 2 \tau \right],
$$
$$
N^{(2)}_{21} = \frac{1}{4} \left[4 \alpha ^2 \rho  \sin \tau +\alpha ^2 \rho  \sin 2 \tau \right],
$$
$$
N^{(2)}_{22} =  \frac{1}{8} \left[ 17 \alpha ^2 \rho ^2-6 \alpha  \beta -16 \alpha ^2 \rho ^2 \cos \tau +8 \alpha  \beta  \cos \tau -\alpha ^2 \rho ^2 \cos 2 \tau  -2 \alpha
\beta  \cos 2\tau \right],
$$
and
$$
N^{(3)}_{11} =
\frac{1}{72} \left[153 \alpha ^3 \rho ^3 \sin
\tau +36 \alpha ^3 \rho ^3 \sin 2 \tau
-5 \alpha ^3
\rho ^3 \sin 3 \tau -72 \alpha ^2 \beta  \rho
\sin \tau -18 \alpha ^2 \beta  \rho  \sin 2 \tau -4
\alpha ^2 \beta  \rho  \sin 3 \tau \right],
$$
$$
N^{(3)}_{12} =  \frac{1}{432} \left[2284 \alpha ^3 \rho ^4-1438
\alpha ^2 \beta  \rho ^2+180 \alpha  \beta^2-2187 \alpha ^3 \rho ^4 \cos \tau  -108 \alpha^3 \rho ^4 \cos 2 \tau+11 \alpha ^3 \rho ^4 \cos
3 \tau
\right.
$$
$$
\left.
+1539 \alpha ^2 \beta  \rho ^2 \cos
\tau -162 \alpha ^2 \beta  \rho ^2 \cos 2 \tau
+61
\alpha ^2 \beta  \rho ^2 \cos 3 \tau -270 \alpha
\beta ^2 \cos \tau+108 \alpha  \beta ^2 \cos 2
\tau -18 \alpha  \beta ^2 \cos 3 \tau \right],
$$
$$
N^{(3)}_{21} = \frac{1}{48} \left[98 \alpha ^3 \rho ^2-20 \alpha
^2 \beta -129 \alpha ^3 \rho ^2 \cos \tau +30
\alpha ^3 \rho ^2 \cos 2 \tau
+\alpha ^3 \rho ^2
\cos 3 \tau+30 \alpha ^2 \beta  \cos \tau -12
\alpha ^2 \beta  \cos 2 \tau+2 \alpha ^2 \beta
\cos 3 \tau \right],
$$
$$
N^{(3)}_{22} =  \frac{1}{144} \left[-531 \alpha ^3 \rho ^3 \sin
\tau +54 \alpha ^3 \rho ^3 \sin 2 \tau+\alpha ^3
\rho ^3 \sin 3 \tau
+144 \alpha ^2 \beta  \rho
\sin \tau +36 \alpha ^2 \beta  \rho  \sin 2 \tau +8
\alpha ^2 \beta  \rho  \sin 3 \tau \right].
$$
\end{widetext}

\end{document}